\documentclass[aps,prb,twocolumn,floatfix,showpacs]{revtex4}
\usepackage{amsmath}
\usepackage{graphicx}

\newcommand{\g}{{\bf g}}
\renewcommand{\r}{{\bf r}}
\renewcommand{\k}{{\bf k}}
\newcommand{\khat}{{\bf \hat{k}}}

\begin{document}

\title{Exchange and correlation as a functional of the
       local density of states}

\author{ Jos\'e M. Soler }
\affiliation{ Departamento de F\'{\i}sica de la Materia Condensada, C-III,
              Universidad Aut\'{o}noma de Madrid,
              E-28049 Madrid, Spain }

\date{\today}

\begin{abstract}
   A functional $E_{xc}[\rho(\r,\epsilon)]$ is presented, in which the 
exchange and correlation energy of an electron gas depends on the 
local density of occupied states. 
   A simple local parametrization scheme is proposed, entirely 
from first principles, based on the decomposition of the 
exchange-correlation hole in scattering states of different 
relative energies.
   In its practical Kohn-Sham-like form, the single-electron orbitals 
become the independent variables, and an explicit formula for the
functional derivative is obtained.
\end{abstract}



\maketitle

\section{Introducction}

   Density functional theory \cite{Hohemberg-Kohn1964,Kohn-Sham1965} 
(DFT) has become an essential tool in the study and simulation of 
molecules and condensed systems.
   Its evolution from the initial local density approximation (LDA) 
to the generalized gradient approximation 
\cite{Becke1988,Lee-Yang-Parr1988,Perdew-Burke-Ernzerhof1996} (GGA),
the mixing of exact exchange \cite{Becke1993,ErnzerhofScuseria1999}
and, more recently, to the meta-GGAs 
\cite{Becke1998,Tao2003} 
has brought a substantial increase in accuracy.
   However, this evolution is not systematic and the holy grail of 
``chemical accuracy'' ($\sim$ 1 Kcal/mole for reaction energies) 
still appears rather distant.
   Perhaps more importantly, basic difficulties remain unsolved,
specially for strongly correlated and magnetic systems, as well as
for excited states.
   Among the latter, the systematic underestimation of the insulating
band gap is particularly troublesome for the simulation of solids.

   In this work, I propose a functional for the exchange and
correlation energy $E_{xc}$, which might potentially overcome some of 
these difficulties.
   Like in the meta-GGAs, its basic ingredient is not just the
electron density.
   Rather, $E_{xc}$ depends on a richer and more informative function,
namely the local density of occupied states.
   Since this function changes much more drastically and transparently 
than the total density for excited states, the functional
can in principle describe these in addition to the ground state.
   Furthermore, the functional form can be rationalized naturally
in terms of electron-electron scattering states, what in principle
allows also to parametrize the functional in an entirely ab initio
and unambiguous way.

\section{Density of states functional}

   Let us first define the local density of occupied kinetic-energy 
states as
\begin{eqnarray}
   \rho(\r,\epsilon) = \sum_{i=1}^N 
      && \int d^{3N}\r~|\Psi(\r_1,...,\r_N)|^2
         ~\delta^3(\r_i-\r) \nonumber \\
      && \times~\delta(\epsilon_i(\r_1,...,\r_N)-\epsilon)
\label{rho_re}
\end{eqnarray}
with
\begin{equation}
   \epsilon_i(\r_1,...,\r_N) = - \frac{\hbar^2}{2 m_e} 
      \Re \frac{\nabla^2_i \Psi(\r_1,...,\r_N)}{\Psi(\r_1,...,\r_N)}
\label{e_i}
\end{equation}
where $\Re$ means real part and I will omit spin degrees of freedom to
simplify the notation.
   Notice that the first two moments of $\rho(\r,\epsilon)$,
\begin{equation}
   \rho(\r) = \int_{-\infty}^{+\infty} d\epsilon~\rho(\r,\epsilon)
\label{rho_r}
\end{equation}
and
\begin{equation}
   \tau(\r) = \int_{-\infty}^{+\infty}
        d\epsilon~\rho(\r,\epsilon)~\epsilon
\label{tau}
\end{equation}
are the electron and the kinetic energy densities.
   The former is the functional variable in all DFT's and
the latter is used as an additional auxiliary function
in meta-GGAs \cite{Becke1998,Tao2003}
(with $\rho(\r)$ and $\tau(\r)$ obtained
from a noninteracting wave function, see below).
   These may thus be considered as particular cases of a
more general functional $E[\rho(\r,\epsilon)]$.

   Notice that, since the wavefunction $\Psi(\r_1,...,\r_N)$ is
determined~\cite{Hohemberg-Kohn1964} by the electron density 
$\rho(\r)$, the density of states
$\rho(\r,\epsilon)$ is also fully determined by $\rho(\r)$,
and therefore it formally does not add any variational freedom.
   However, in practice $\rho(\r,\epsilon)$ provides explicitly
much more information on the electronic state than $\rho(\r)$,
and therefore it allows for a more accurate parametrization of
the exchange and correlation energies.
   On this respect, $\rho(\r,\epsilon)$ is a reasonable half way
in complexity between $\rho(\r)$ and $\Psi(\r_1,...,\r_N)$.
   Thus, I express the total energy as
\begin{eqnarray}
   E[\rho(\r,\epsilon)]
     & = & T[\rho(\r,\epsilon)] + U_H[\rho(\r)]
           + E_{xc}[\rho(\r,\epsilon)] \nonumber \\
     & + & \int d^3\r~\rho(\r)~v_{ext}(\r)
\label{Esplit}
\end{eqnarray}
where $T$ is the kinetic energy
\begin{equation}
   T[\rho(\r,\epsilon)] = \int d^3\r~\tau(\r),
\label{T}
\end{equation}
$U_H$ is the Hartree energy
\begin{equation}
   U_H[\rho(\r)] = \int\int d^3\r~d^3\r' 
                   ~\frac{e^2 \rho(\r) \rho(\r')}{|\r-\r'|}
\label{UH}
\end{equation}
$v_{ext}(\r)$ is the external potential,
and $E_{xc}[\rho(\r,\epsilon)]$ is some approximation to the exchange 
and correlation energy, which includes everything else.

   Although Eq.~(\ref{T}) is in principle the total kinetic energy,
in practice we will use a density
$\rho(\r,\epsilon)$ obtained from a single-determinant wave function
$\Psi_0$ and therefore $E_{xc}[\rho(\r,\epsilon)]$ must also contain the
difference in kinetic energy between $\Psi_0$ and the exact wave function.
   More precisely, let us introduce another functional
\begin{eqnarray}
   E^0[\rho^0(\r,\epsilon)] 
     & = & T[\rho^0(\r,\epsilon)] + U_H[\rho(\r)]
           + E^0_{xc}[\rho^0(\r,\epsilon)] \nonumber \\
     & + & \int d^3\r~\rho(\r)~v_{ext}(\r)
\label{E0}
\end{eqnarray}
where
\begin{equation}
   \rho^0(\r,\epsilon) = \sum_{\alpha=1}^N 
      |\psi_{\alpha}(\r)|^2~\delta(\epsilon^0_\alpha(\r)-\epsilon),
\label{rho0_re}
\end{equation}
\begin{equation}
   \epsilon^0_\alpha(\r) = - \frac{\hbar^2}{2 m_e} 
      \Re \frac{\nabla^2 \psi_\alpha(\r)}{\psi_\alpha(\r)},
\label{e0_i}
\end{equation}
and $\psi_\alpha(\r)$ are single-electron, orthonormal, occupied orbitals.

   For a given density $\rho(\r)$, the functions 
$\rho(\r,\epsilon)$ and $\rho^0(\r,\epsilon)$ are different.
   Furthermore, since the same density $\rho(\r)$ can result from
different sets of $\psi_\alpha(\r)$'s, and therefore from different
functions $\rho^0(\r,\epsilon)$, there is an intrinsic indetermination
in the functional $E^0$.
   An example of such indetermination is a rotation in the space
of occupied orbitals, which leaves invariant $\rho(\r)$ but not
$\rho^0(\r,\epsilon)$.
   The indetermination can be removed by specifying an explicit
way of obtaining $\psi_\alpha(\r)$ from $\rho(\r)$.
   A natural choice is to minimize $T[\rho^0(\r,\epsilon)]$,
under the constraint (\ref{rho_r}).
   This leads to the Kohn-Sham equations
\begin{equation}
\left( - \frac{\hbar^2}{2 m_e} \nabla^2 + v(\r) - 
       \epsilon_{\alpha} \right) \psi_{\alpha}(\r) = 0
\label{Schroedinger}
\end{equation}
where $v(\r)$ is the Lagrange multiplier associated to the 
constraint (\ref{rho_r}), and $\epsilon_{\alpha}$ is that associated 
to the normalization constraint for $\psi_{\alpha}$.
   However, I will simply consider the indetermination in 
$\rho^0(\r,\epsilon)$ as an additional freedom to achieve a
practical and accurate parametrization of $E^0[\rho^0(\r,\epsilon)]$.

   In Eqs.~(\ref{Esplit}) and (\ref{E0}), $E[\rho(\r,\epsilon)]$ 
and $E^0[\rho^0(\r,\epsilon)]$ are formally assumed 
to be infinite for all functions, $\rho(\r,\epsilon)$ and
$\rho^0(\r,\epsilon)$, which cannot be expressed in the forms 
(\ref{rho_re}) and (\ref{rho0_re}), respectively.
   In practice, this condition needs not be dealt with, since we will
use the standard Kohn-Sham procedure \cite{Kohn-Sham1965} of 
minimizing the total energy as a function of $\psi_{\alpha}(\r)$ 
rather than of $\rho^0(\r,\epsilon)$.
   In what follows, I will refer only to $E^0_{xc}$, and 
$\rho^0(\r,\epsilon)$, but I will drop the superscript 0 to simplify 
the notation.

   The advantage of $\rho(\r,\epsilon)$ as a functional variable is that 
it provides much more explicit information on the system state than 
$\rho(\r)$.
   Thus, it naturally leads to an energy and orbital-dependent 
exchange-correlation potential, much like quasiparticle self energies, 
thus offering hope of improving the description of excited states and
band gaps.
   Also, since the density of states is very different in localized
and extended systems, it offers the possibility to address 
electron localization and self-interaction problems.
   In a different context, the relevance of spectral functions in
the description of electron localization and strong correlation
has been stressed by dynamical mean field theory~\cite{Georges1996}.
   Finally, the success of meta-GGA functionals in improving 
atomization energies gives us some confidence that the proposed 
functional may also improve considerably on formation and 
reaction energies.
   On the other hand, it might be suspected that, since the local 
density of states is much richer in structure than the total density, 
a functional dependent on it should be also much harder to parametrize.
   I will show that, at least in principle, this is not the case, and 
that unambiguous and physically rooted parametrizations of $E_{xc}$ 
are possible.

\section{Functional form}

   Here I propose the local functional form \cite{emass}
\begin{equation}
   E_{xc} = \int d^3\r \int\int_{-\infty}^{+\infty} 
       d\epsilon~d\epsilon'~K_{xc}(\rho(\r),\epsilon,\epsilon') 
       ~\rho(\r,\epsilon)~\rho(\r,\epsilon')
\label{Exc}
\end{equation}
where the kernel $K_{xc}$ is an average of the interaction energy 
between two electrons with wave vectors $\k$ of fixed size 
$k = \sqrt{2 m_ e \epsilon}/\hbar$ but random direction $\khat$:
\begin{eqnarray}
   K_{xc}(\rho,\epsilon,\epsilon') 
      & = & \frac{1}{2} \int\int 
            \frac{d^2 \khat}{4 \pi}~\frac{d^2 \khat'}{4 \pi}
            ~e_{xc}(\rho,|\k-\k'|) \nonumber \\
      & = & \frac{1}{2} \int dq ~e_{xc}(\rho,q) ~P(q|\epsilon,\epsilon')
\label{Kxc}
\end{eqnarray}
where the factor 1/2 corrects double counting and 
$P(q|\epsilon,\epsilon')$ is the probability that the two electrons, 
with kinetic energies $\epsilon = \hbar^2 k^2 / 2m_e$ and 
$\epsilon' = \hbar^2 k'^2 / 2m_e$, have relative momentum 
$\hbar q = \hbar |\k - \k'|/2$:
\begin{eqnarray}
   P(q|\epsilon,\epsilon') 
     &=& \int\int \frac{d^2 \khat}{4 \pi}~\frac{d^2 \khat'}{4 \pi}
         ~\delta\left(\frac{|\k-\k'|}{2} - q\right) \nonumber \\
     &=& \left\{ 
           \begin{array}{ll} 
             2q/(k k')  & \mbox{if $|k-k'| < 2q < |k+k'|$} \\
             0                & \mbox{otherwise.}
           \end{array}
         \right.
\label{Pqee}
\end{eqnarray}
   $e_{xc}(\rho,q)$ is the energy change of an electron pair, with
relative momentum $\hbar q$, in a homogeneous gas of density $\rho$
(assumed unpolarized for simplicity),
when their antisymmetry is imposed and their interaction potential 
$e^2/r$ is switched on. 
   It may be written as $e_{xc}(\rho,q) = e_x(q) + e_c(\rho,q)$.
   Expanding the plane waves in spherical harmonics
\begin{eqnarray}
   e_x(q) 
     &=& \sum_{l=0}^\infty (2l+1) ~(s_l - 1) \int_0^\infty dr
         ~4 \pi r^2 \frac{e^2}{r} ~j_l^2(qr) \nonumber \\
     &=& -\frac{\pi}{2q^2}
\label{ex}
\end{eqnarray}
where the factor $s_l = 1 - (-1)^l/2$ accounts for the antisymmetry of 
equal-spin electron-pair wavefunctions~\cite{GoriGiorgi-Perdew2001}.
   $e_c$ may be approximated by
\begin{eqnarray}
   && e_c(\rho,q) = \sum_{l=0}^\infty (2l+1) ~s_l
      \int_0^\infty dr ~4 \pi r^2 \nonumber \\
   && \times \left[ \phi_l(\rho,q,r) ~H_l ~\phi_l(\rho,q,r)
                   - j_l(qr) ~H_l ~j_l(qr) \right],
\label{ec}
\end{eqnarray}
where $H_l = T_l + e^2/r$ and
\begin{equation}
   T_l \equiv -\frac{\hbar^2}{2 \mu r}~\frac{\partial^2}{\partial r^2} r +
         \frac{\hbar^2 l(l+1)}{2 \mu r^2},
\label{Tl}
\end{equation}
with $\mu = m_e/2$.
   $\phi_l(\rho,q,r)$ and $j_l(qr)$ are the interacting and noninteracting
scattering wavefunctions:
\begin{equation}
  \left( T_l + V_{int}(\rho,r)
   - \frac{\hbar^2 q^2}{2 \mu} \right) \phi_l(\rho,q,r) = 0,
\label{phiofq}
\end{equation}
and $j_l(qr)$ is the solution of the same equation with $V_{int} = 0$,
i.e.\ a spherical Bessel function.
   $V_{int}(\rho,r)$ is an effective interaction potential for the 
screened Coulomb repulsion, designed to reproduce accurately the pair
correlation function of the homogeneous electron gas~\cite{Vint}:
\begin{equation}
   g(\rho,r) = \sum_{l=0}^\infty (2l+1)~s_l \int_0^{k_F} dq~P(q|\rho)
               ~\phi_l^2(\rho,q,r),
\label{gofr}
\end{equation}
where $P(q|\rho)$ is the probability of finding an electron pair with
relative momentum $\hbar q$ in a homogeneous gas of density $\rho$. 
   For $q < k_F = (3 \pi^2 \rho)^{1/3}$:
\begin{eqnarray}
   P(q|\rho) 
     &=& \int\int_0^{\epsilon_F} d\epsilon~d\epsilon'~P(q|\epsilon,\epsilon')
         ~P(\epsilon|\rho) ~P(\epsilon'|\rho) \nonumber \\
     &=& 24 \frac{q^2}{k_F^3} - 36 \frac{q^3}{k_F^4} + 12 \frac{q^5}{k_F^6}
\label{Pqrho}
\end{eqnarray}
where $P(\epsilon|\rho) = 3k/k_F^3$ is proportional to the free-electron 
density of states.
   Eqs.(\ref{phiofq}-\ref{gofr}) define a problem which is presently under 
very active research \cite{Overhauser1995,GoriGiorgi-Perdew2001,
Davoudi2002:Hart,Davoudi2002:func,Nagy2003,Ziesche2003,Corona2003}.
   Furthermore, there a other possible alternatives for $e_c$ in 
Eq.~(\ref{ec}), like using $e^2/r$ instead of $H_l$ and integrating in
the coupling constant of the electron-electron interaction.
   Therefore, the actual parametrization of $V_{int}(\rho,r)$ and 
$e_{xc}(\rho,q)$ will be left for a future work.

   A problem with the above approach is that, while the local kinetic energies 
$\epsilon_\alpha(\r)$ are negative in classically-forbidden regions, they are 
always positive in a homogeneous gas.
   This problem may be overcome by using the Airy gas proposed by
Kohn and Mattsson \cite{Kohn-Mattsson1998} in which the total one-electron
effective potential is a constant-force ramp $v(\r)=-{\bf f}\cdot(\r-\r_0)$.
   Since a constant force does not affect the relative motion of two equal
particles, it is a good approximation to use the same $e_{xc}$ of 
the homogeneous gas and to modify only the probability $P$ 
of equation~(\ref{Kxc}) for the Airy gas.
   To do this, we can use the one-to-one relationship \cite{Kohn-Mattsson1998}
$(\rho(\r),\nabla\rho(\r)) \Leftrightarrow ({\bf f},\r_0)$, making
$P$ and $K_{xc}$ dependent on $|\nabla\rho|$ in addition to $\rho$.
   In fact, this procedure can be used also to derive GGAs and
meta-GGAs from $e_{xc}(\rho,q)$.

\section{Functional derivative}

   To minimize the total energy or to perform ab initio molecular dynamics, we 
need the functional derivative $\delta E_{xc}/\delta\psi^*_\alpha(\r)$ 
\cite{Payne1992:RMP,Car-Parrinello1985}.
   To calculate it, it is convenient to assume that the occupied 
orbitals $\psi_\alpha(\r)$, density $\rho(\r)$, and density of states 
$\rho(\r,\epsilon)$ are given in a mesh of points $\r_i$ and discrete 
energies $\epsilon$.
   Their gradients and laplacians can then be calculated as
\begin{equation}
   \nabla \rho_i \equiv \nabla \rho(\r_i) = \sum_j {\bf g}_{ij} \rho_j
\label{grho}
\end{equation}
\begin{equation}
   \nabla^2 \psi_{i\alpha} \equiv \nabla^2 \psi_\alpha(\r_i) 
      = \sum_j L_{ij} \psi_{j\alpha}
\label{lpsi}
\end{equation}
where $\rho_i \equiv \rho(\r_i), \psi_{j\alpha} \equiv \psi_{\alpha}(\r_j)$,
and ${\bf g}_{ij}, L_{ij}$ are coefficients of some unspecified
finite-difference formulas.
   We can then replace the functional derivatives by ordinary partial
derivatives and recover the continuous limit at the end.
   Using atomic units $(\hbar=e=m_e=1)$ and omitting for brevity the 
grid increments, $\Delta x^3$ and $\Delta\epsilon$, when replacing the 
integrals $\int d^3\r$ and $\int d\epsilon$ by grid sums, the discretized 
equations are
\begin{equation}
   E_{xc} = \sum_j \sum_{\epsilon \epsilon'}
       K_{j \epsilon \epsilon'} \rho_{j \epsilon} \rho_{j \epsilon'}
\label{Excd}
\end{equation}
where 
$K_{j\epsilon\epsilon'} \equiv K_{xc}(\rho(\r_j),|\nabla\rho(\r_j)|,\epsilon,\epsilon')$,
\begin{equation}
   \rho_{j\epsilon} = \sum_\alpha \psi^*_{j\alpha} \psi_{j\alpha} 
       \delta_{j\alpha\epsilon},
\label{rhoje}
\end{equation}
with $\delta_{j\alpha\epsilon}$ a discretized delta function
(remember that $\epsilon$ are now discrete energies) like
\begin{equation}
    \delta_{j\alpha\epsilon} = 
        \delta(\epsilon_{j\alpha} - \epsilon) = 
        \left\{ 
           \begin{array}{ll} \frac{1}{\Delta \epsilon} \left( 
             1 - \frac{|\epsilon_{j\alpha}-\epsilon|}{\Delta\epsilon} \right)
               & \mbox{if $|\epsilon_{j\alpha}-\epsilon| < \Delta\epsilon$} \\
             0 & \mbox{otherwise,}
           \end{array}
        \right.
\label{djae}
\end{equation}
which must verify that 
$\sum_{\epsilon} \delta_{j \alpha \epsilon}=1 ~~\forall \epsilon_{j \alpha}$,
where
\begin{equation}
   \epsilon_{j\alpha} = - \Re \frac{\nabla^2\psi_{j\alpha}}{2\psi_{j\alpha}}
      = -\frac{1}{4} \sum_k L_{jk}
             \left( \frac{\psi_{k\alpha}}{\psi_{j\alpha}} +
                    \frac{\psi^*_{k\alpha}}{\psi^*_{j\alpha}} \right).
\label{eja}
\end{equation}
   I will now consider formally $\psi^*_{j\alpha}$ and $\psi_{j\alpha}$ as 
independent variables to perform the derivatives
\begin{equation}
   \frac{\partial\rho_j}{\partial\psi^*_{i\alpha}} = \delta_{ij} \psi_{j\alpha},
\label{drhodpia}
\end{equation}
where $\delta_{ij}$ is the usual Kronecker delta,
\begin{equation}
   \frac{\partial|\nabla\rho_j|}{\partial\psi^*_{i\alpha}} = 
   \frac{\nabla\rho_j \g_{ji}}{|\nabla\rho_j|} \psi_{i\alpha},
\label{dgrhodpia}
\end{equation}
\begin{equation}
   \frac{\partial\epsilon_{j\alpha}}{\partial\psi^*_{i\alpha}} = 
      -\frac{L_{ji}}{4\psi^*_{j\alpha}} + \frac{\delta_{ij}}{4\psi^{*2}_{j\alpha}}
        \sum_k L_{jk} \psi^*_{k\alpha},
\label{dejadpia}
\end{equation}
\begin{eqnarray}
   \frac{\partial\rho_{j\epsilon}}{\partial\psi^*_{i\alpha}} 
     & = & \delta_{ij} \psi_{j\alpha} \delta_{j\alpha\epsilon}
         + \psi^*_{j\alpha} \psi_{j\alpha} 
           \frac{d\delta_{j\alpha\epsilon}}{d\epsilon_{j\alpha}}
           \frac{\partial\epsilon_{j\alpha}}{\partial\psi^*_{i\alpha}} 
           \nonumber \\
     & = & \left( 
             \delta_{ij} \delta_{j\alpha\epsilon}
           - \frac{L_{ij}}{4}  
             \frac{d\delta_{j\alpha\epsilon}}{d\epsilon_{j\alpha}}
             \right. \nonumber \\
     &   & + \left. 
             \frac{d\delta_{j\alpha\epsilon}}{d\epsilon_{j\alpha}}
             \frac{\delta_{ij}}{4\psi^*_{j\alpha}}
             \sum_k L_{jk} \psi^*_{k\alpha}
           \right) \psi_{j\alpha},
\label{drhojedpia}
\end{eqnarray}
and
\begin{eqnarray}
   \frac{\partial E_{xc}}{\partial\psi^*_{i\alpha}}
     & = & \sum_{j\epsilon\epsilon'}
           \left(
             \frac{\partial K_{j\epsilon\epsilon'}}{\partial\rho_j}
             \frac{\partial\rho_j}{\partial\psi^*_{i\alpha}}
           + \frac{\partial K_{j\epsilon\epsilon'}}{\partial|\nabla\rho_j|}
             \frac{\partial|\nabla\rho_j|}{\partial\psi^*_{i\alpha}}
           \right) \rho_{j\epsilon} \rho_{j\epsilon'}
           \nonumber \\
    &    & + 2 \sum_{j\epsilon\epsilon'} K_{j\epsilon\epsilon'} \rho_{j\epsilon'}
             \frac{\partial\rho_{j\epsilon}}{\partial\psi^*_{i\alpha}}.
\label{dExcdpia_1}
\end{eqnarray}
   Substituting (\ref{drhodpia}), (\ref{dgrhodpia}), and (\ref{drhojedpia}) into 
(\ref{dExcdpia_1}) we obtain
\begin{eqnarray}
   \frac{\partial E_{xc}}{\partial\psi^*_{i\alpha}}
     & = & \left( u_i + 2\sum_\epsilon \delta_{i\alpha\epsilon} v_{i\epsilon} \right)
           \psi_{i\alpha}
           \nonumber \\
     &   & + \frac{\psi_{i\alpha}}{2\psi^*_{i\alpha}} \sum_j L_{ij} \psi^*_{j\alpha}
             \sum_\epsilon \frac{d\delta_{i\alpha\epsilon}}
                                {d\epsilon_{i\alpha}} v_{i\epsilon} 
           \nonumber \\
     &   & - \frac{1}{2} \sum_j L_{ij} \psi_{j\alpha} \sum_\epsilon
             \frac{d\delta_{j\alpha\epsilon}}{d\epsilon_{j\alpha}}
             v_{j\epsilon}.
\label{dExcdpia_2}
\end{eqnarray}
   I have defined
\begin{equation}
   u_i  \equiv \sum_{\epsilon\epsilon'} 
                  \frac{\partial K_{j\epsilon\epsilon'}}{\partial\rho_i}
                  \rho_{i\epsilon} \rho_{i\epsilon'}
                - \sum_j \frac{\g_{ij}\nabla\rho_j}{|\nabla\rho_j|}
                  \sum_{\epsilon\epsilon'}
                  \frac{\partial K_{j\epsilon\epsilon'}}{\partial|\nabla\rho_j|}
                  \rho_{j\epsilon} \rho_{j\epsilon'}
\label{ui}
\end{equation}
\begin{equation}
   v_{i \epsilon} \equiv \sum_{\epsilon'} K_{i\epsilon\epsilon'} \rho_{i\epsilon'}
\label{vi}
\end{equation}
where I have used that $L_{ji} = L_{ij}$ and $\g_{ji} = -\g_{ij}$.
   Returning to the continuous limit,
\begin{eqnarray}
   \frac{\delta E_{xc}}{\delta\psi^*_{\alpha}(\r)}
     && = \left( u(\r) + 2\int d\epsilon 
                  ~\delta(\epsilon_\alpha(\r)-\epsilon) ~v(\r,\epsilon)
          \right) \psi_\alpha(\r)
          \nonumber \\
     && + \frac{\psi_\alpha(\r)\nabla^2\psi^*_\alpha(\r)}{2\psi^*_\alpha(\r)}
          \int d\epsilon 
                 \frac{d\delta(\epsilon_\alpha(\r)-\epsilon)}
                      {d\epsilon_\alpha(\r)} v(\r,\epsilon)
          \nonumber \\
     && - \frac{1}{2} \nabla^2 
          \left( \psi_\alpha(\r) \int d\epsilon
                 \frac{d\delta(\epsilon_\alpha(\r)-\epsilon)}
                      {d\epsilon_\alpha(\r)} v(\r,\epsilon)
             \right)
\label{dExcdpar}
\end{eqnarray}
\begin{eqnarray}
  && u(\r) = \int\int d\epsilon d\epsilon' 
            \frac{\partial K_{xc}(\r,\epsilon,\epsilon')}{\partial\rho(\r)} 
            \rho(\r,\epsilon) \rho(\r,\epsilon') 
            \nonumber \\
        & & - \nabla 
            \left( 
              \frac{\nabla\rho(\r)}{|\nabla\rho(\r)|}
           \right. \nonumber \\
        & & \left.
              \int\int d\epsilon d\epsilon'
              \frac{\partial K_{xc}(\r,\epsilon,\epsilon')}
                   {\partial|\nabla\rho(\r)|} 
              \rho(\r,\epsilon) \rho(\r,\epsilon')
            \right),
\label{ur}
\end{eqnarray}
\begin{equation}
  v(\r,\epsilon) = \int d\epsilon'~K_{xc}(\r,\epsilon,\epsilon') 
                                  ~\rho(\r,\epsilon')
\label{vr}
\end{equation}
where I am using the shorthand notation 
$K_{xc}(\r,\epsilon,\epsilon') \equiv 
K_{xc}(\rho(\r),|\nabla\rho(\r)|,\epsilon,\epsilon')$.
   Using integration by parts in (\ref{dExcdpar}) to integrate out the 
delta functions
\begin{eqnarray}
   \frac{\delta E_{xc}}{\delta\psi^*_{\alpha}(\r)}
     &=& \left( u(\r) + 2 v(\r,\epsilon_\alpha(\r)) \right) \psi_\alpha(\r)
          \nonumber \\
     && + \frac{\psi_\alpha(\r)\nabla^2\psi^*_\alpha(\r)}{2\psi^*_\alpha(\r)}
                ~\frac{\partial v(\r,\epsilon_\alpha(\r))}
                      {\partial\epsilon_\alpha(\r)}
          \nonumber \\
     && - \frac{1}{2} \nabla^2 
          \left( \psi_\alpha(\r) \frac{\partial v(\r,\epsilon_\alpha(\r))}
                                      {\partial\epsilon_\alpha(\r)}
          \right).
\label{dExcdpar_2}
\end{eqnarray}
   In practice, $v(\r,\epsilon_\alpha(\r))$ and its derivative will be 
interpolated from the values at a number of discrete energies, i.e.
\ $v(\r_i,\epsilon_\alpha(\r_i)) = 
\sum_\epsilon v_{i\epsilon} c_\epsilon(\epsilon_\alpha(\r_i))$,
where the interpolation functions $c_\epsilon$ are determined by,
e.g.\ Lagrange or spline interpolation formulas.
   Comparison with Eq.~(\ref{dExcdpia_2}) shows that the discrete delta
values $\delta_{i\alpha\epsilon}$, used to distribute the
contribution of state $\alpha$ to $\rho(\r_i,\epsilon)$, at the
discrete energies $\epsilon$, are simply 
$\delta_{i\alpha\epsilon} = c_\epsilon(\epsilon_\alpha(\r_i))$.
   Thus, for example, Eq.~(\ref{djae}) corresponds to a simple linear 
interpolation for $v(\r_i,\epsilon_\alpha(\r_i))$.
   Though Eqs.~(\ref{ur}-\ref{dExcdpar_2}) are considerably more complicated 
than the functional derivatives in LDA or GGA functionals, the gradients and 
laplacians can be calculated efficiently using fast Fourier transforms
or finite differences, at a quite affordable computational cost.

\section{Conclusions}

   In conclusion, I have proposed a functional form for the exchange
and correlation energy, with the potential to describe more accurately 
excitation energies and band gaps than the present forms.
   The basic ideas, embodied in Eq.~(\ref{Exc}), are that the 
interaction between two electrons depends on their relative motion, 
and that the exchange and correlation energy of an electron depends 
on its kinetic energy.
   These ideas are physically transparent and intuitively appealing, and
therefore they may be expected to allow for significant improvements
in the description of the inhomogeneous electron gas.

\begin{acknowledgments}
   I want to acknowledge useful conversations with Pablo 
Garc\'{\i}a-Gonz\'alez and Pedro Tarazona. 
   This work has been supported by the Fundaci\'on Ram\'on Areces and by
Spain's MCyT grants BFM2000-1312 and BFM2003-03372.
\end{acknowledgments}

\bibliographystyle{apsrev}
\bibliography{dft,hole,dsft}

\begin{thebibliography}{22}
\bibitem[{Vin()}]{Vint}
\bibinfo{note}{In a more elaborate formulation, $V_{int}$ will depend also on
  the relative spins of the two electrons, as well as on the electron gas
  polarization.}
\end{thebibliography}

\end{document}